\documentclass{pasj01}
\Received{$\langle$reception date$\rangle$}
\Accepted{$\langle$acception date$\rangle$}
\Published{$\langle$publication date$\rangle$}

\usepackage{multirow}
\usepackage{color}

\begin{document}

\title{Detection of the hard X-ray non-thermal emission from Kepler’s supernova remnant}
\author{%
  Tsutomu \textsc{Nagayoshi},\altaffilmark{1}
  Aya \textsc{Bamba},\altaffilmark{2,3}
  Satoru \textsc{Katsuda},\altaffilmark{1}
  and
  Yukikatsu \textsc{Terada}\altaffilmark{1}}
\altaffiltext{1}{Graduate School of Science and Engineering, Saitama University, Shimo-Okubo 255, Sakura, Saitama 338-8570}
\email{nagayoshi@heal.phy.saitama-u.ac.jp}
\altaffiltext{2}{Department of Physics, The University of Tokyo, 7-3-1 Hongo, Bunkyo, Tokyo 113-0033, Japan}
\altaffiltext{3}{Research Center for the Early Universe, School of Science, The University of Tokyo, 7-3-1 Hongo, Bunkyo-ku, Tokyo 113-0033, Japan}

\maketitle

\begin{abstract}
  We report the first robust detection of the hard X-ray emission in the 15--30\,keV band from Kepler's supernova remnant with the silicon PIN-type semiconductor detector of the hard X-ray detector (HXD-PIN) onboard Suzaku.
  The detection significance is 7.17 $\sigma$  for the emission from  Kepler's entire X-ray  emitting region.
  The energy spectrum is found to be well reproduced by a single power-law  function with a photon index of $3.13^{+1.85+0.69}_{-1.52-0.36}$,
  where the first and second errors represent  90\%-statistical and systematic errors, respectively.
  The X-ray flux  is determined to be 2.75$_{-0.77-0.82}^{+0.78+0.81}\times10^{-12}$\,erg\,s$^{-1}$\,cm$^{-2}$ in the 15--30\,keV band.
  The wider-band X-ray spectrum in the 3--30\,keV band, where the soft X-ray Suzaku/XIS spectrum is combined, shows that the non-thermal component does not have a significant X-ray roll-off structure.
  We find that the  broad-band energy spectrum  from the radio band, X-ray data of this work, and TeV upper limits can be reproduced with the one-zone leptonic model with  a
  roll-off energy of $\nu_{\mathrm{roll}}=1.0\times10^{17}$\,Hz and  magnetic field strength of $B>40\,\mathrm{\mu G}$.
  Application of the diagnostic method using indices in the soft and hard X-ray band to the data indicates that
  the maximum energy of the accelerated electrons in Kepler's SNR is limited by the age of the remnant.
  The indication is consistent with the results of the one-zone leptonic modeling.
\end{abstract}

\section{Introduction}\label{sec:introduction}

Supernova remnants (SNRs) are widely considered to be the primary origin of the galactic cosmic rays (GCRs), which are highly accelerated particles with energies up to $\sim10^{15}$\,eV.
The most plausible acceleration mechanism at SNRs  is diffusive shock acceleration (DSA)
(e.g., \cite{1977ICRC...11..132A,1978MNRAS.182..147B,1978ApJ...221L..29B}).
The key feature of this process is that the acceleration is first order in the shock velocity and automatically results in a power-law spectrum with energy spectral index of $\sim$2.
 However Monte Carlo simulations show that steeper particle spectra $>2$ is required to be consistent with the number of expected SNRs to be seen in TeV gamma-ray and the detected TeV SNRs \citep{2013MNRAS.434.2748C}.
In addition, the basic DSA is not able to  explain  some of the observed characteristics, including  the maximum energy of the particles,  configuration of the magnetic field, and injection efficiency for the particle acceleration.
More advanced, non-linear DSA models have been studied to address these issues (e.g., \cite{2002A&A...395..943B,2003A&A...412L..11B,2010ApJ...718...31P}).
The models predict that the magnetic field strength  in the vicinity of the SNR shock is amplified to $\sim100\,\mathrm{\mu G}$,  with which particles can be accelerated up to $\sim10^{15}$\,eV.
The scenario is supported by the  detection of  a high magnetic field strength in X-ray\ observations (e.g., \cite{2007Natur.449..576U,2008ApJ...677L.105U,2018ApJ...868L..21B,2020ApJ...894...50O,2020PASJ..tmp..234M}).
\par
Since \citet{1995Natur.378..255K} discovered the first observational evidence that electrons are accelerated up to multi-TeV energies in SN\,1006,
many astronomers have followed suit and made extensive study of the relations between GCRs and SNRs, both observationally and theoretically.
In particular, observations in the X-ray energy band  often provide invaluable information.
In the X-ray band, young SNRs emit synchrotron radiation,  typically  with rather steep spectral indices of $\Gamma$=2--3.5. This type of X-ray spectra suggest that they are emitted by high-energy electrons that have  a rather steep  energy distribution \citep{2012A&ARv..20...49V}. The highest energy of the electrons should be
  close to the maximum accelerated electron energy $E_{\mathrm{max,e}}$, which is determined from
 the magnetic field strength and  shock speed \citep{1999ApJ...525..368R}.
In addition, the curvature of an X-ray synchrotron spectrum  provides information of the particle acceleration process, such as the so-called age-limited case and loss-limited case \citep{2014RAA....14..165Y}.
Thus, study of  X-ray spectra of SNRs helps  us gain insight about the particle acceleration and  local environments.
\par
Kepler's SNR is the remnant of SN\,1604 and  one of the youngest SNRs in our Galaxy.
It has a roughly spherical shell with a diameter of $\sim200''$, accompanied with two protrusions in the north-west and south-east, and has been observed extensively in the  radio and X-ray  bands (e.g., \cite{1988ApJ...330..254D,2002ApJ...580..914D,2007ApJ...668L.135R}). 
The center of the remnant is located at the Galactic coordinates $l=4.5^{\circ}$ and $b=6.8^{\circ}$. 
The distance to the remnant is still under discussion.
\citet{2005AdSpR..35.1027S}  reported a distance of 3.9$\pm$1.4\,kpc from their proper motion measurement.
 Almost contradictorily, \citet{1999AJ....118..926R} estimated it to be 4.8--6.4\,kpc, using H\emissiontype{I} data obtained with the VLA.
 Their estimated distance  has been supported by  TeV gamma-ray observations.
Based on the upper limit in the TeV gamma-ray range, the distance is estimated to be at least 6.4\,kpc, where the typical type Ia SN explosion model \citep{2008A&A...488..219A} was employed. 
 To summarize conservatively, the distance to the remnant is  3--7\,kpc \citep{Kerzendorf2014}.
We adopt a distance of 4\,kpc throughout this paper.
\par
In the soft X-ray band below 10\,keV,  non-thermal emission from Kepler's SNR  has been detected by  past X-ray missions, which gave a
 roll-off frequency of $\nu_{\mathrm{roll}}$=1.1--7.9$\times10^{17}$\,Hz. This value suggests  $E_{\mathrm{max,e}}\sim$50--130\,TeV  in the standard synchrotron radiation model emitted by  accelerated electrons  in a uniform magnetic field of $\sim10\,\mathrm{\mu G}$
\citep{2005ApJ...621..793B,2004A&A...414..545C,2008ApJ...689..225K,1999PASJ...51..239K}.
Given that the radial profile of the non-thermal X-rays from Kepler's SNR follows the thin-filament structures at the shell, non-linear DSA may be playing a key role in electron acceleration.
\citet{2005ApJ...621..793B} extracted  spatially-resolved X-ray energy spectra  from the thin-filament structures of the shock front and estimated its
roll-off frequency  to be $\nu_{\mathrm{roll}}=3.6^{+3.3}_{-1.6}\times10^{17}$\,Hz.
\citet{2004A&A...414..545C}  made image and radial profile analyses using {\it XMM-Newton} and
found the X-ray emission in the south-east region  to be largely non-thermal.
\citet{2007ApJ...668L.135R} also
confirmed this feature with
the deep {\it Chandra} observation and found that a few regions in Kepler's SNR were dominated  with  a continuum component, which  is possibly synchrotron emission.
Indeed, the measured shock speeds $v_{\mathrm{s}}$  from various regions of Kepler's SNR have been consistently higher  than the minimum value required to emit X-ray synchrotron emission $\sim2,000\,\mathrm{km\,s^{-1}}$ \citep{1999A&A...351..330A}, the fact of which supports the synchrotron-origin hypothesis; $v_{\mathrm{s}}$ from
the X-ray brightest knots was estimated to be 9,100--10,400\,km\,s$^{-1}$ \citep{2017ApJ...845..167S}, that from the ejecta  distributed widely over
the inner of the rim  was 2,000--3,000\,km\,s$^{-1}$ \citep{2018PASJ...70...88K},
and that from the overall rims of the remnant was 2,000--4,000\,km\,s$^{-1}$  \citep{2008ApJ...689..225K}.
As such, several soft X-ray observations claimed to have found  evidence that the non-thermal X-ray emission from Kepler's SNR  originates  in  accelerated electrons at the shell.
\par
\indent However, in the hard X-ray band above 10 keV, the flux and spectral shape of the synchrotron radiation have not been determined.
In this paper, we report the first  robust detection of the wide-band non-thermal spectrum of Kepler's SNR taken with  Suzaku \citep{2007PASJ...59S...1M}.
The Suzaku observations and data reduction of Kepler's SNR are described in section\,2,  our spectral analysis, in section\,3, and  discussion and summary, in section\,4.

\section{Observations and data reduction}\label{sec:dataset}
The Suzaku satellite has  two types of instruments,  the X-ray imaging spectrometers (XISs: \cite{2007PASJ...59S..23K})
and  hard X-ray detector (HXD: \cite{2007PASJ...59S..35T}), covering the soft (0.2--10.0\,keV) and hard (10--600\,keV) X-ray energy bands, respectively.
The HXD is a well-type-phoswitch scintillation counters, whose main photo-absorbers consist of silicon PIN-type semiconductor detectors (HXD-PIN) and Gd$_2$SiO$_5$ (hereafter GSO) crystal scintillators (HXD-GSO),
covering energy  bands of 10--70\,keV and 40--600\,keV, respectively.
In this  work, we analyzed the HXD-PIN data.
No significant detection was made in the HXD-GSO data.
Regarding the XIS data, we  adopted the spectrum presented in \citet{2015ApJ...808...49K}, which is taken from the observation with  an observation ID of 505092040.
\par
Suzaku  made in total eight  observations of Kepler's  SNR region in 2010 September, 2010 October, 2011 February, and 2011 March. The total exposure time is 568.94\,ks.  Table\,\ref{tbl:observation} summarizes the details of the Suzaku observations.
\begin{table*}
  \tbl{ Suzaku observations used in this  work.
    The observation IDs 502078010 and 505092040 are used for the analyses in this paper.
    The NXB uncertainties  for those two observation IDs are consistent with \citet{2009PASJ...61S..17F} within  1$\sigma$ of the statistical error.}
      {%
        \begin{tabular}{ccccccccc}
          \hline\hline
          \multirow{2}{*}{ID} & \multirow{2}{*}{Date} & RA & Dec & \multicolumn{2}{c}{Earth occultation Observation} & \multicolumn{3}{c}{On Source observation}  \\
          & & [deg] & [deg] & Exposure\footnotemark[$*$] [ks] & Uncertainties [\%] & Exposure\footnotemark[$*$] [ks] & $a$ [count sec$^{-1}$] & $b$ \\
          \hline\noalign{\vskip3pt}
          502078010 & 2008-02-18 & 262.66 & -21.54 & 52.81 &  1.0$\pm$1.1   &  98.95 & 0.04$\pm$0.01 & 0.94$\pm$0.05 \\
          505092010 & 2010-09-30 & 262.67 & -21.44 &  0.25 & 16.3$\pm$17.5  &  17.85 & 0.06$\pm$0.03 & 0.88$\pm$0.11 \\
          505092020 & 2010-10-06 & 262.67 & -21.44 & 33.77 &  0.1$\pm$1.5   &  99.80 & 0.07$\pm$0.01 & 0.86$\pm$0.06 \\
          505092030 & 2011-02-23 & 262.65 & -21.53 & 20.50 &  2.9$\pm$1.9   &  29.10 & 0.04$\pm$0.02 & 0.97$\pm$0.08 \\
          505092040 & 2011-02-28 & 262.66 & -21.54 & 90.86 &  0.2$\pm$0.9   & 113.73 & 0.04$\pm$0.01 & 0.97$\pm$0.04 \\
          505092050 & 2011-03-08 & 262.66 & -21.54 & 70.29 &  0.8$\pm$0.9   & 122.47 & 0.21$\pm$0.01 & 0.35$\pm$0.04 \\
          505092060 & 2011-03-14 & 262.66 & -21.54 &  8.32 &  1.4$\pm$2.6   &  38.61 & 0.01$\pm$0.03 & 1.04$\pm$0.10 \\
          505092070 & 2011-03-29 & 262.66 & -21.54 & 54.73 &  1.4$\pm$1.2   & 112.20 & 0.07$\pm$0.01 & 0.83$\pm$0.05 \\
          \hline\hline
        \end{tabular}
      }
      \label{tbl:observation}
      \begin{tabnote}
        \hangindent6pt\noindent
        \hbox to6pt{\footnotemark[$*$]\hss}\unskip: The exposure are the ones after processing of the HXD-PIN.
      \end{tabnote}
\end{table*}
\par
We used the software package {\tt Heasoft 6.25} with {\tt CALDB 2019-09-13} for the general analysis.
The data were reprocessed and screened in  the standard  procedure, using the Suzaku reprocessing tool {\tt aepipeline}.
We applied the standard criteria 
\footnote{({\tt T\_SAA\_HXD>500 and TN\_SAA\_HXD>180}) and ({\tt HXD\_HV\_W[0123]\_CAL>700 and HXD\_HV\_T[0123]\_CAL>700}) and ({\tt AOCU\_HK\_CNT3\_NML\_P==1 and ANG\_DIST<1.5}); see the reference of \texttt{Heasoft} for detail.}  to filter the data.
We also applied the following filtering criteria: Earth elevation angle (ELV) $>$5$^{\circ}$ and geomagnetic cut-off rigidity (COR) $>$ 6\,GV for  the ``On Source'' data,
and ELV $<$ $-$5$^{\circ}$ and COR $>$ 8\,GV for the Earth-occultation data, which are used to evaluate the systematic uncertainties of the Non-X-ray Background (NXB) model and data selection.
\par
The NXB for the HXD is estimated using the methods described in \citet{2009PASJ...61S..17F}.
We used the NXB events of  {\tt LCFITDT} (bgd-d)  of version  2.0ver0804.
The uncertainties of the bgd-d NXB model  was reported to be 1--3\% by \citet{2009PASJ...61S..17F}.
The cosmic X-ray background (CXB) was calculated  with an assumption of a power-law continuum model with an exponential cut-off at 40 keV, as previously determined  with the {\it HEAO-1 A2} data \citep{1987IAUS..124..611B}.
We used {\tt hxdpinxbpi} to extract the  spectra for the source region, NXB, and CXB.
\par
The combined XIS spectrum by \citet{2015ApJ...808...49K}  that we adopted
were accumulated from only the front-illuminated CCDs (XIS0 and XIS3) because they have a slightly better spectral resolution than the back-illuminated CCD (XIS1).
The data reduction for the XIS data  were carried out  with the same  criteria applied to  the HXD-PIN data.
In addition, we calculated the effective area for the XIS by using {\tt xissimarfgen} \citep{2007PASJ...59S.113I},
to which we put {\it Chandra}'s image in 0.3--1.7\,keV to take into account the spatial distribution of Kepler's SNR.
\section{Spectral analysis}\label{sec:analysis}
Spectral analysis was performed using {\sc XSPEC}\footnote{https://heasarc.gsfc.nasa.gov/xanadu/xspec/} v12.10.1.
We used Cash statistic C \citep{1979ApJ...228..939C,2017A&A...605A..51K}, which allows background subtraction in {\sc XSPEC}  by means of the W-statistic, for model-fitting of the spectral data.
\subsection{Hard X-ray spectrum with the HXD-PIN}
Before spectral analysis of the HXD-PIN, we evaluated the reproducibility of the NXB model (bgd-d) for the observation data.
First, we calculated the reproducibility of the bgd-d model
by comparing count rates in the 15--40\,keV band between the spectra during the Earth occultations based on the criteria described in section\,\ref{sec:dataset} and NXB model.
The reproducibility of the bgd-d model was derived to be in the range of 0.1–16\% (the sixth column in table\,\ref{tbl:observation}).
Since the  uncertainty in the NXB modeling is  larger in  shorter observations \citep{2009PASJ...61S..17F}, 
we  excluded the two shorter-exposure observations of IDs of 505092010 and 505092060  in our analysis.
In addition, since the HXD-PIN sensitivity is determined primarily by the systematic error below $\sim$30\,keV,
we also  excluded the observation ID 505092030,  which has an uncertainty  of 2.9\%, to minimize the uncertainty of  the obtained spectral shape.
\par
\begin{figure*}
  \begin{center}
    \includegraphics[width=40mm]{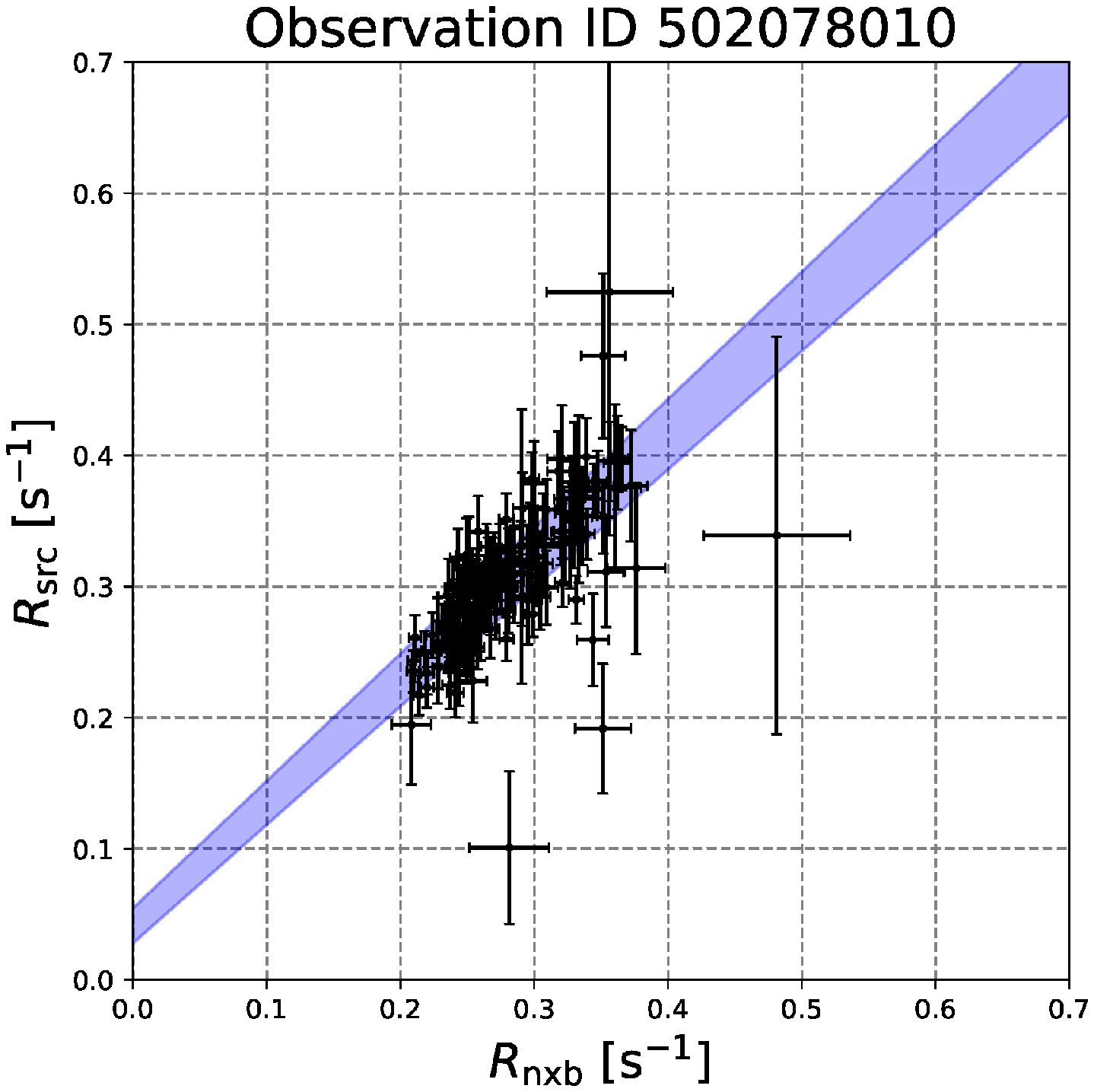}
    \includegraphics[width=40mm]{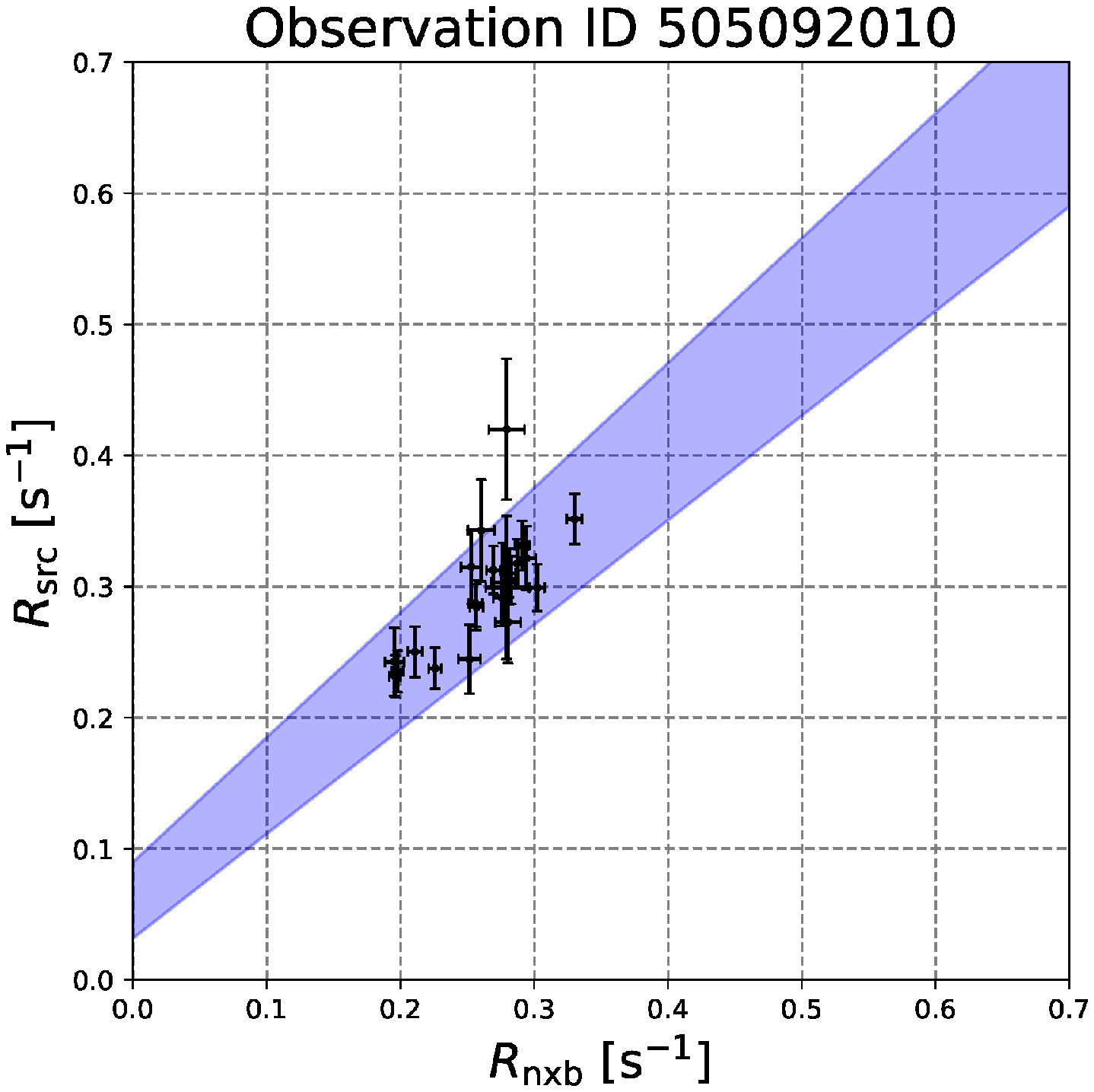}
    \includegraphics[width=40mm]{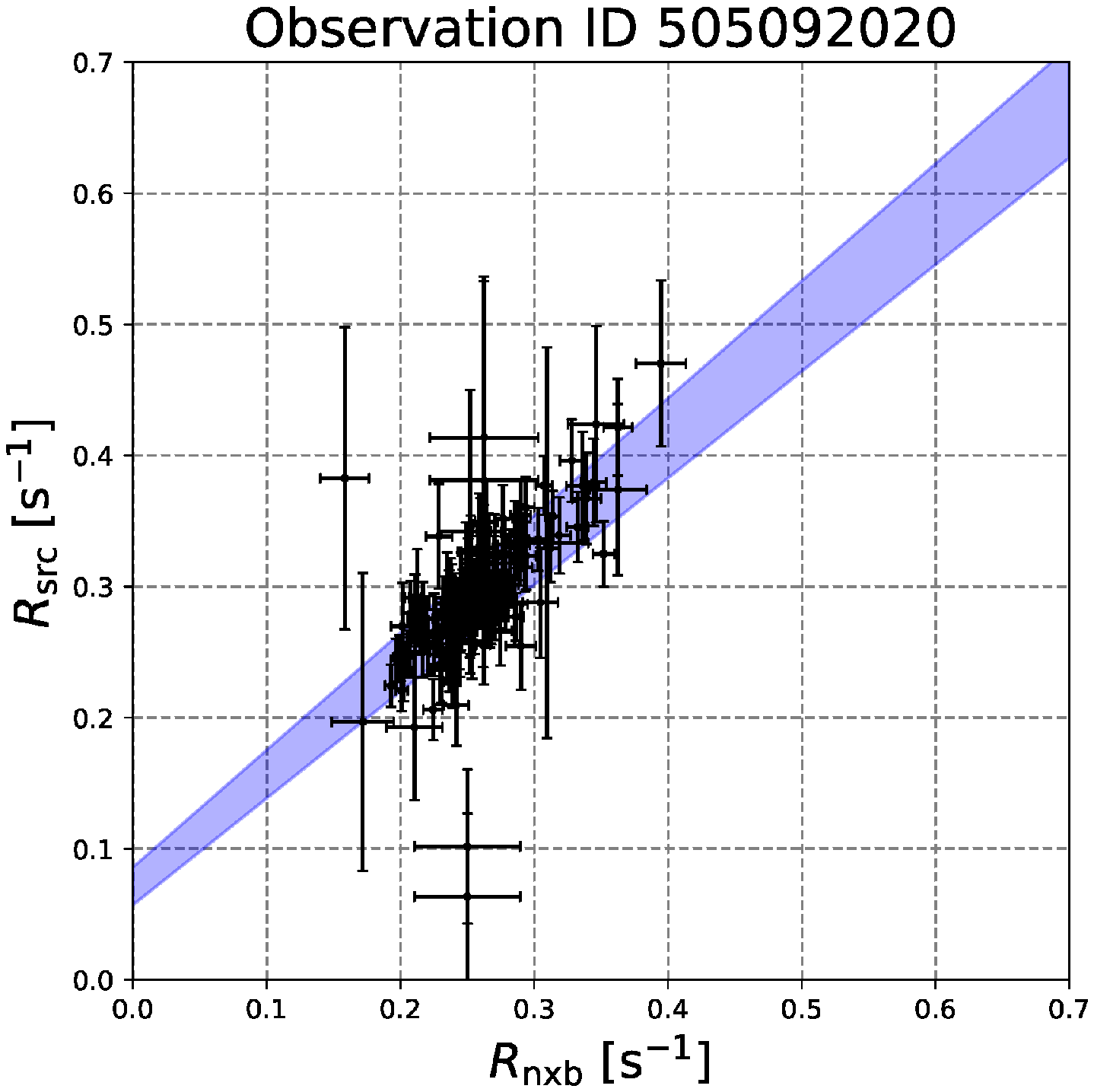}
    \includegraphics[width=40mm]{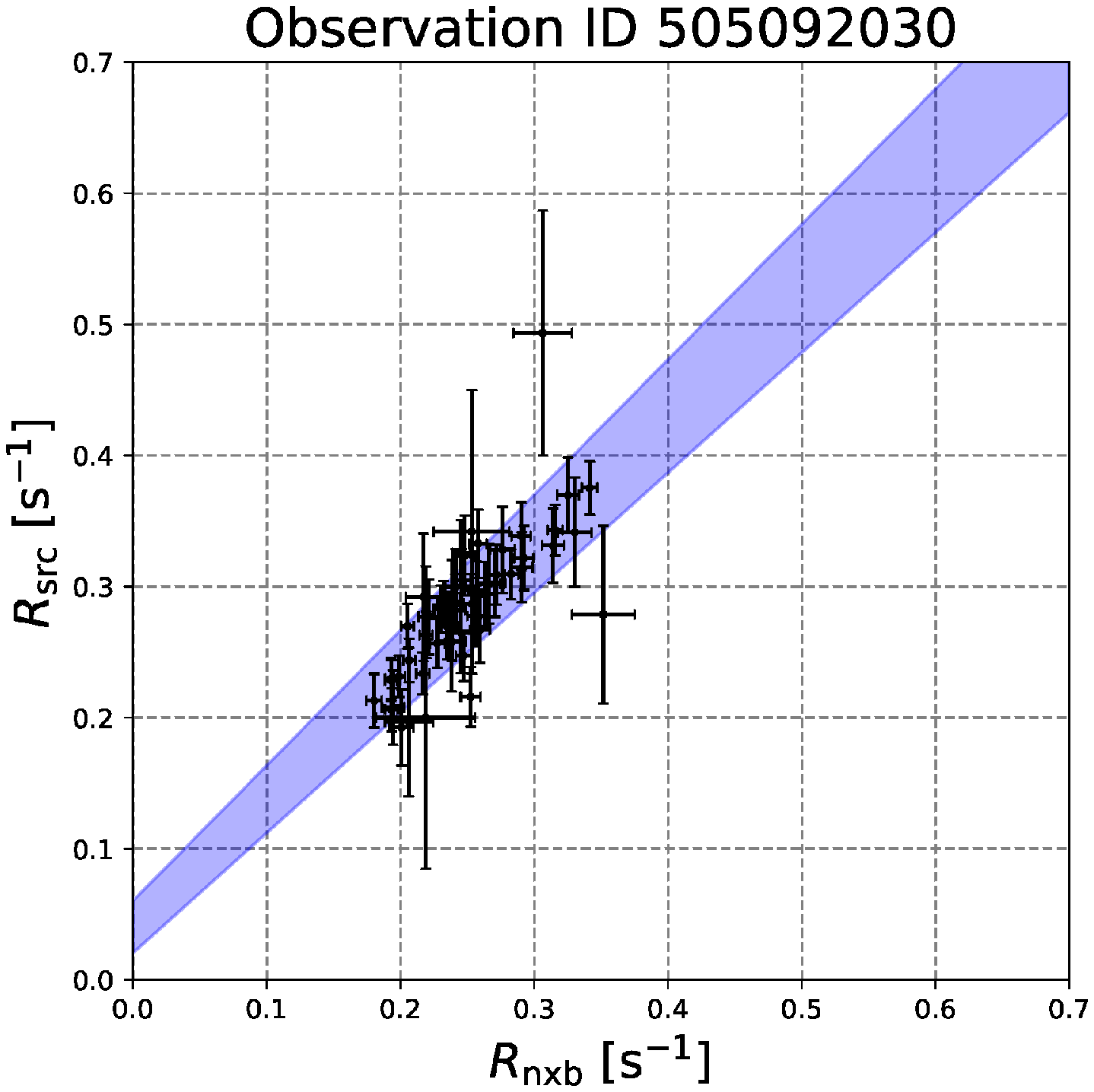}\\
    \includegraphics[width=40mm]{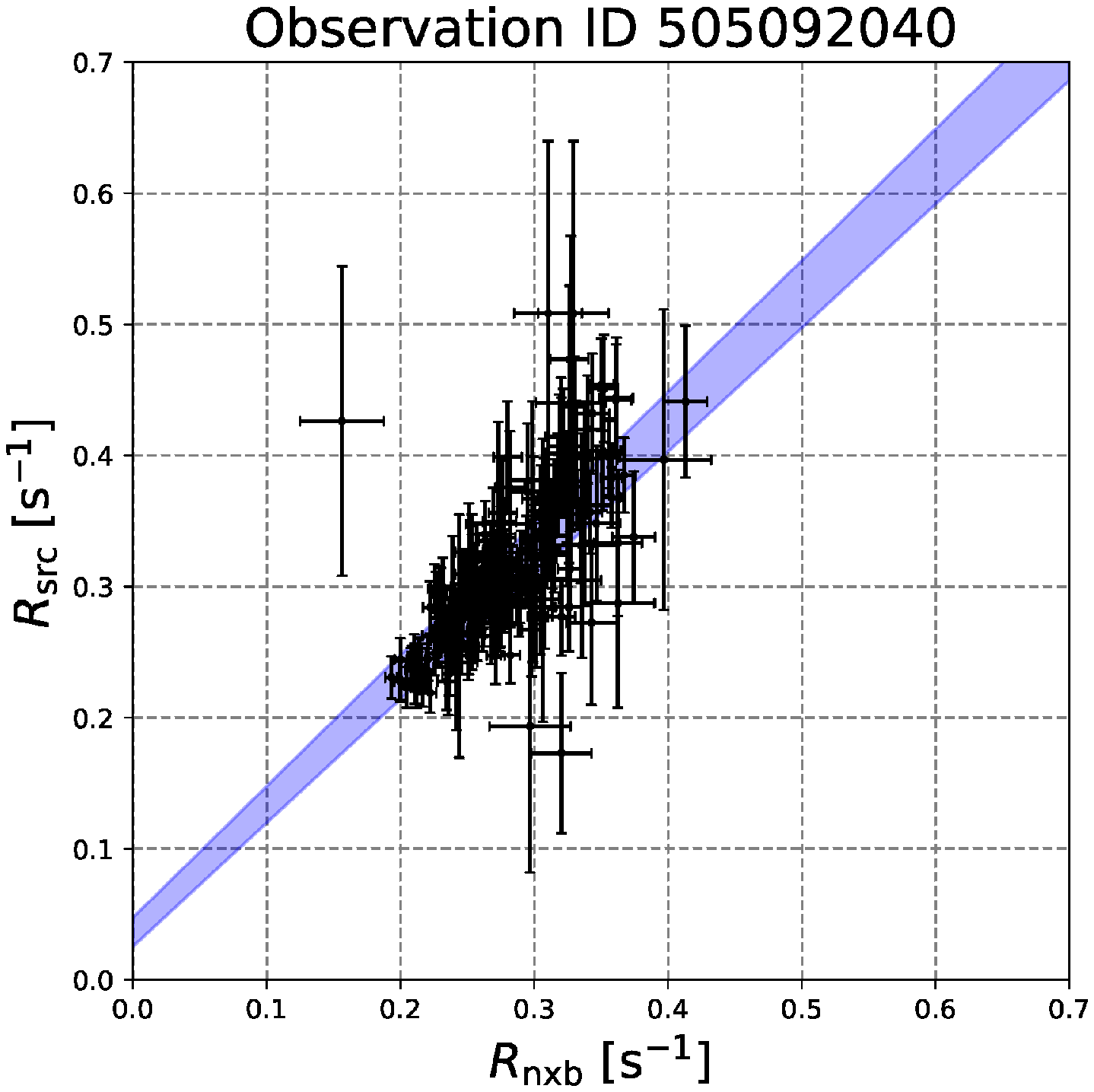}
    \includegraphics[width=40mm]{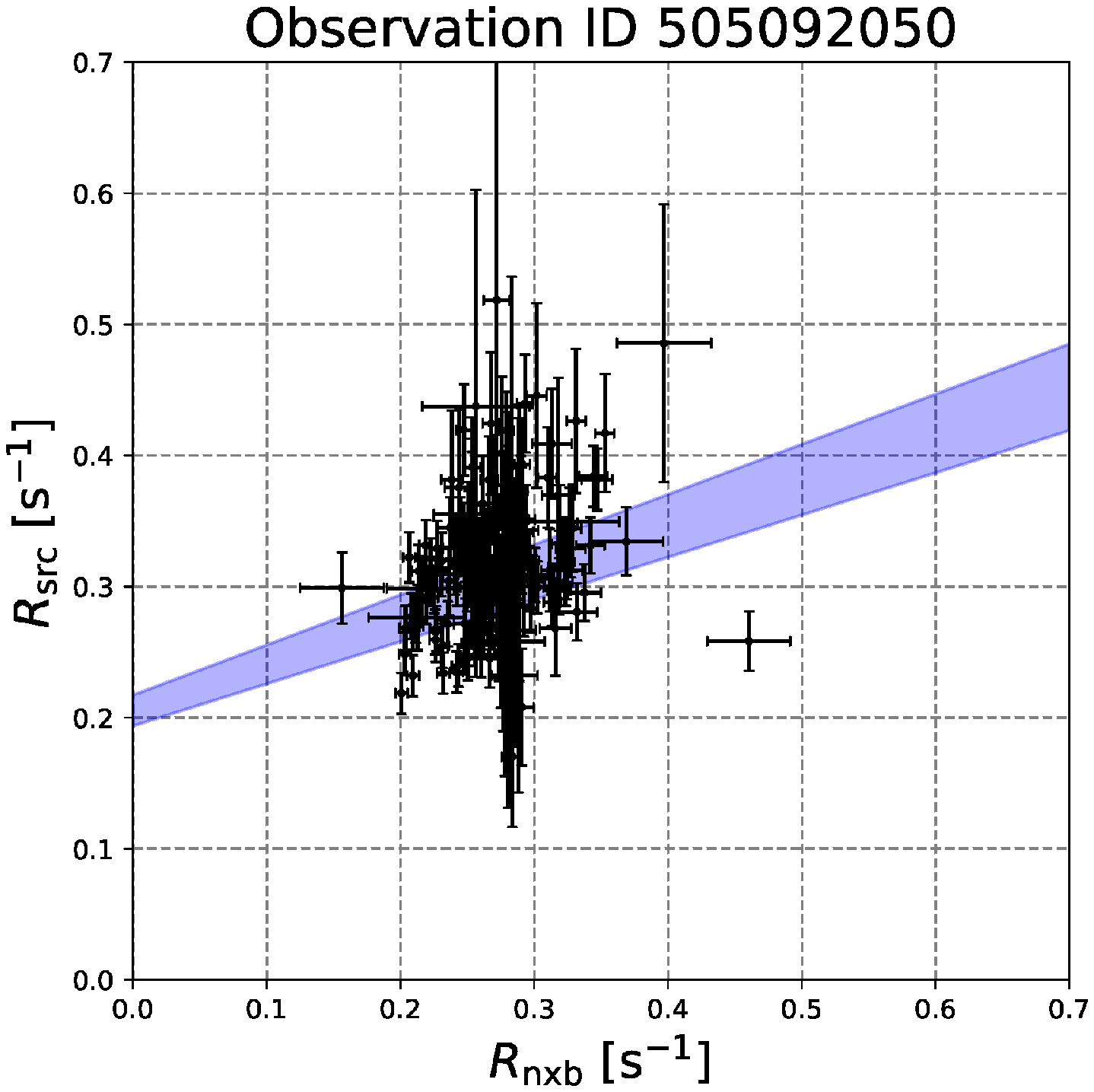}
    \includegraphics[width=40mm]{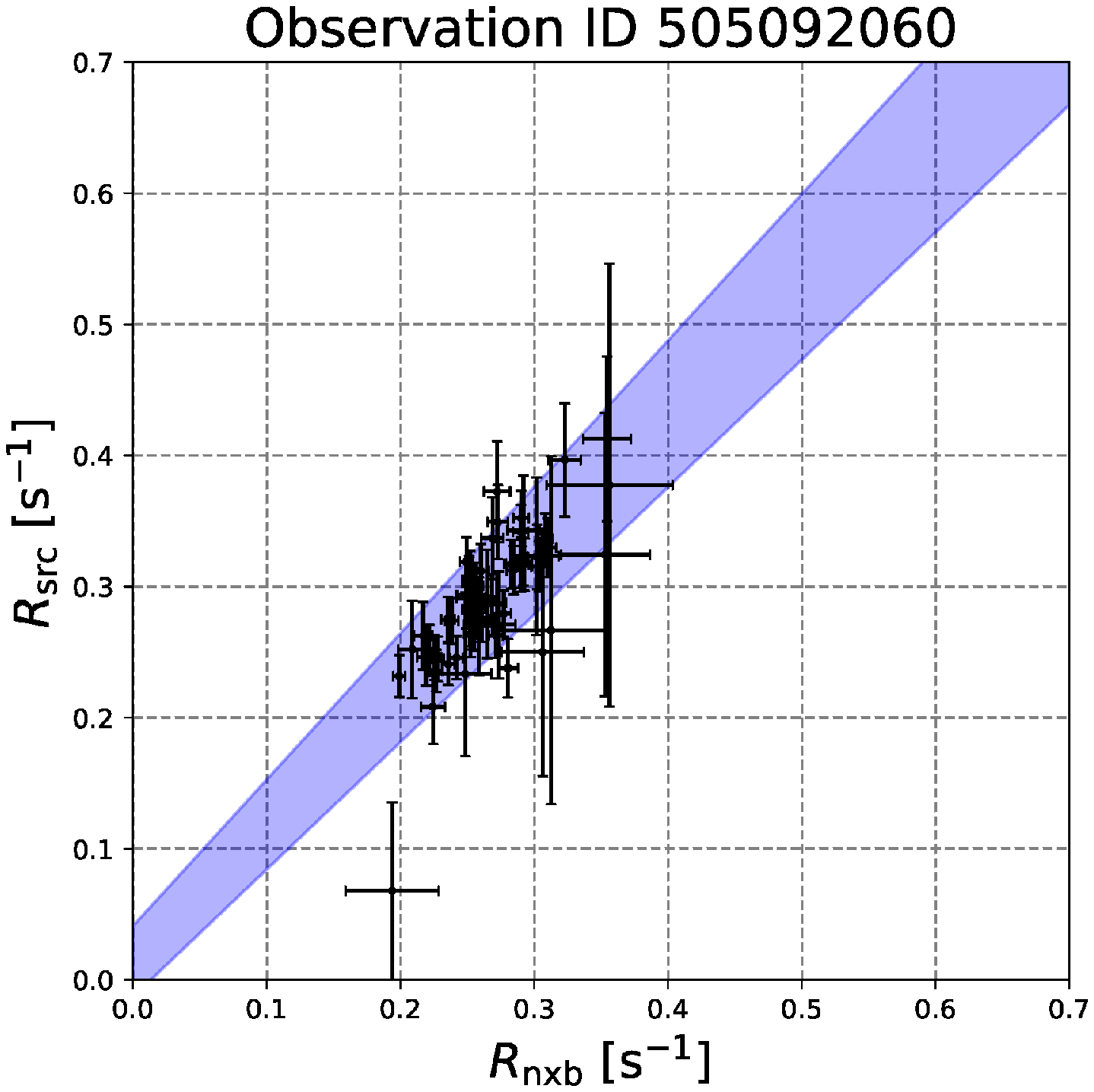}
    \includegraphics[width=40mm]{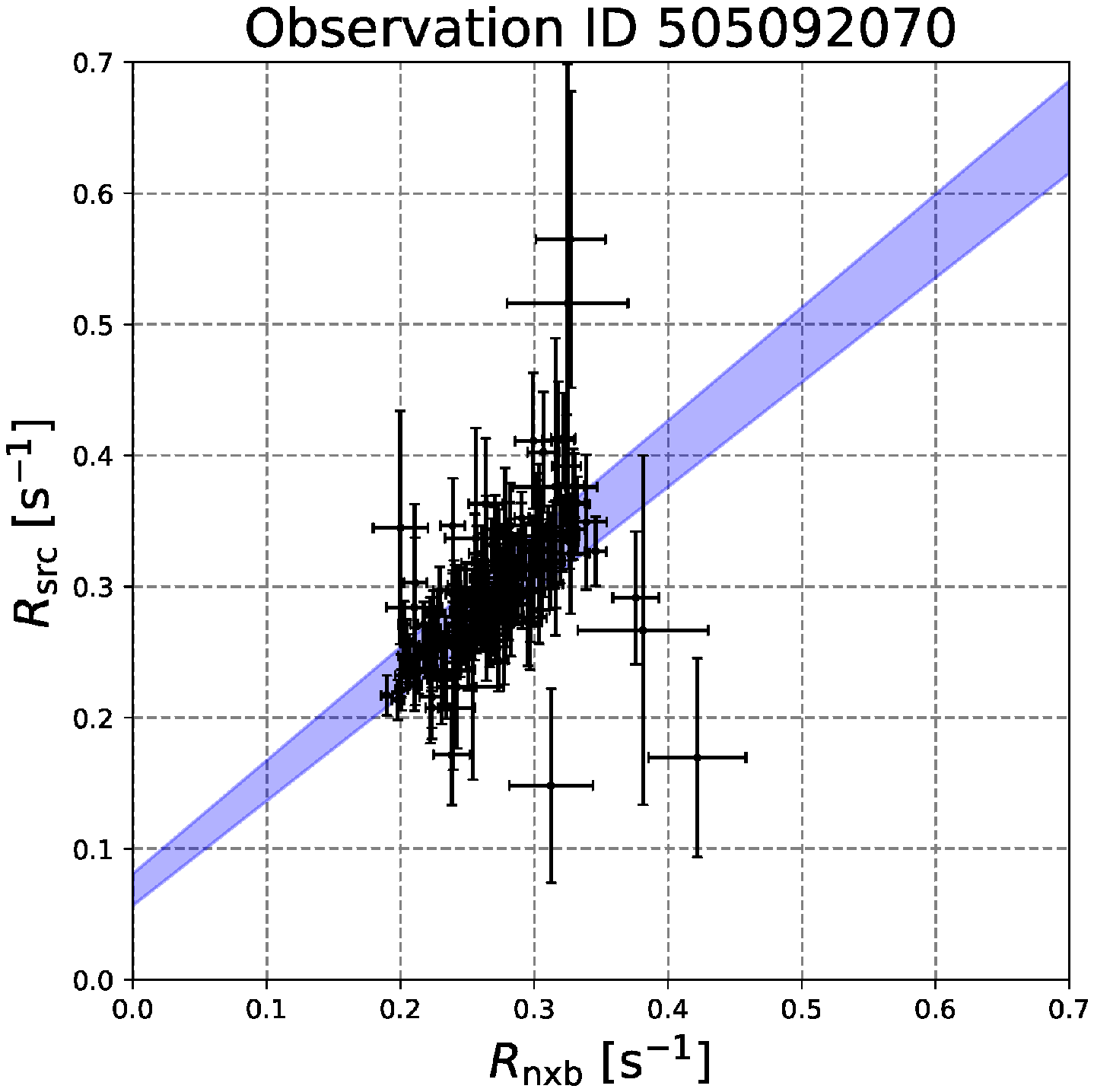}
  \end{center}
  \caption{Comparison of the NXB model count-rate (horizontal axis) and HXD-PIN data count-rate (vertical axis) in the 15--40\,keV band.
    Light-blue shaded areas show the regions of 1-$\sigma$  errors. }
  \label{fig:hxd_pin_nxb_model}
\end{figure*}

Next, we evaluated the effect of poorly  reproduced  NXB models
by inspecting the count-rate correlation between  the ``On Source'' data (see section\,\ref{sec:dataset})
and the NXB model in the same period in the 15--40\,keV band.
We split the total observation period of the  ``On Source'' data in each observation  into 1\,ks
and compared  the NXB count-rate between the data and NXB model in 15--40\,keV.
The correlation plots were fitted  with  a function,
\begin{equation}
  R_{\mathrm{src}}=a+bR_{\mathrm{nxb}},
  \label{eq:rsrc_rnxb}
\end{equation}
where $R_{\mathrm{src}}$ and $R_{\mathrm{nxb}}$ are the observed X-ray  and predicted NXB count-rates, respectively,
$a$ is the X-ray emission such as the CXB and/or that from astronomical objects, and $b$ is
the slope of the relation between the observation data and the NXB model
which should be unity in the ideal case.
Figure\,\ref{fig:hxd_pin_nxb_model} shows the data and fitting results, and the rightmost columns of  table\,\ref{tbl:observation} lists the determined parameters $a$ and $b$.
\par
\par
 The NXB was found to be relatively poorly reproduced in three of the correlation plots (observations IDs of 505092020, 505092050, and 505092070), in addition to those of the already excluded data, 
in the high-count rate range of each ``On source'' observation,  where the parameter $b$ of  these observation IDs is not close to 1.
Indeed,  a hint of this  tendency  had been already reported by \citet{2008PASJ...60S.153B}.
Therefore, we  excluded the observation IDs 505092020, 505092050, and 505092070 with  a criterion of $0.90<b<1.10$ from further analyses.
Consequently,  we analyzed the HXD-PIN data of the observation IDs 502078010 and 505092040 in this work and the total exposure of  the data set is 143.76\,ks.
We adopted a systematic error of 1\% for the following spectral analyses based on the NXB model uncertainties for observation IDs 502078010 and 505092040.
\par
Since the emission region of Kepler's SNR is slightly extended (the angular size of $\sim$200$''$),
we take into account the angular response of the HXD-PIN (the so-called ``arf'' in {\sc XSPEC}) in estimating the effective area.
Specifically, the effective area ratio of a point source to the spatial structure of Kepler's SNR needs to be calculated.
Note that the  angular response of the PIN detectors has a pyramidal shape \citep{2007PASJ...59S..53K,teradaieee2005}. 
For simplicity, we assumed that the  area of the hard X-ray emission was the same as that of the soft X-ray emission, and  that both had flat spatial distributions.
Then, we calculated effective area ratio of the pyramidal shape response to the response for point sources to be 0.96.
This factor is  incorporated in the flux estimation described in the later sections in this paper.
\par
\begin{figure}
  \begin{center}
    \includegraphics[width=85mm]{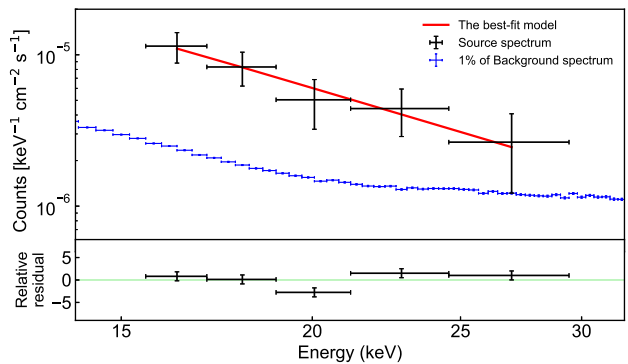}
  \end{center}
  \caption{HXD-PIN spectrum of Kepler's SNR in the 15--30\,keV band. 
    Upper panel shows the background-subtracted HXD-PIN spectrum (black crosses),
     best-fit single power-law model (red line) and  background spectrum (blue line) scaled to 1\% of its intensity.
    Lower panel shows the relative residuals between the data points and  best-fit model.
    The  size of the error bars corresponds to the 68\,\% confidence range  under an assumption of Poisson distribution.}
  \label{fig:spec_pin}
\end{figure}

Figure\,\ref{fig:spec_pin} shows the merged PIN spectrum (the observation IDs  502078010 and 505092040)
 where the NXB and CXB are subtracted,  overlaid with the best-fit power-law model, together with  the background spectrum scaled to 1\% of its intensity for  comparison.
The detection significance  in an energy band of 15.6--29.4\,keV  was 7.17$\sigma$.
The spectrum  was found to be well reproduced by a single power-law  function with c-stat/dof=0.12.
The best-fit photon index  was 3.13$_{-1.52-0.36}^{+1.85+0.69}$ and the flux  in the 15--30\,keV band is
2.75$_{-0.77-0.82}^{+0.78+0.81}\times10^{-12}\,\mathrm{erg}\,\mathrm{cm}^{-2}\,\mathrm{s}^{-1}$ (table\,\ref{tbl:param_fit_pin}), where
the first and second errors represent 90\%-statistical  and systematic errors, the latter of which  originated from  the  NXB model uncertainties of 1\%.

\subsection{Broad-band X-ray spectra with the XIS and HXD-PIN}\label{subsec:broadfit}
The thermal emission  that dominates the soft X-ray energy band  may somewhat contaminate the hard X-ray emission in the HXD-PIN data.
We use the soft X-ray spectrum obtained by \citet{2015ApJ...808...49K}.
\citet{2015ApJ...808...49K}
showed that the soft X-ray spectrum was well reproduced with thermal models,
with their analysis of the combined spatially-integrated spectra measured
with the {\it XMM-Newton} Reflection Grating Spectrometer (RGS) \citep{denHerder2001} and {\it Chandra} ACIS \citep{Garmire2003} for the energy range below 2\,keV and Suzaku XIS (using only the front-illuminated CCDs, XIS0 and XIS3) for  above 2\,keV.
\citet{2015ApJ...808...49K} fitted these spectra in an energy range of 0.4--7.5\,keV with a model consisting of an absorbed, {\tt vpshock} (shock plasma model) + three {\tt vnei}s (non-equillibrium ionization thermal plasma model) + {\tt power-law} + several Gaussian components  with the XSPEC package \citep{Arnaud1996}.
The {\tt vpshock}, three {\tt vnei}, and {\tt power-law} components in the model correspond to the emissions from the circumstellar medium (CSM),
SN ejecta (Ejecta 1: the lower-temperaturee component, Ejecta 2: the higher-temperature component, Ejecta 3: the Fe-rich component),
and synchrotron radiation, respectively.
The several Gaussian components represent emission lines of Fe L and/or Ne K, Cr K, and Mn K \citep{2015ApJ...808...49K}.
\par
\begin{figure}
  \begin{center}
    \includegraphics[width=85mm]{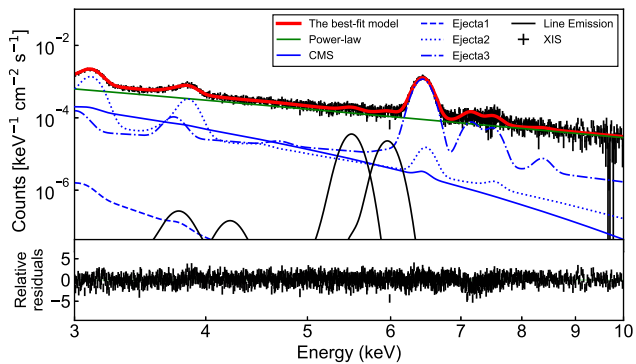}
  \end{center}
  \caption{XIS spectrum of Kepler's SNR in the 3--10\,keV band. 
    Upper panel shows the background-subtracted XIS spectrum taken from \citet{2015ApJ...808...49K}  with black crosses and  best-fit model (see text) with a red line. Lower panel shows the relative residuals between the data points and  best-fit model. 
    The  size of the error bars corresponds to the 68\,\% confidence range  under an assumption of Poisson distribution.  
  }
  \label{fig:spec_xis}
\end{figure}

In our fitting analyses, we applied the model presented in \citet{2015ApJ...808...49K} for thermal components but with the latest AtomDB ver 3.0.9\footnote{http://www.atomdb.org}.
The difference of the thermal parameters between those by \citet{2015ApJ...808...49K} and our work is summarized in appendix~\ref{sec:app_a}.
Then we applied the obtained new best-fit model to the XIS spectrum in the 3--10 \,keV band, allowing only the normalization parameter of each thermal model.
In this fitting, the normalization for the CSM and Ejecta 1 are fixed because the contributions of these components in the 3–-10\,keV band were relatively minor contribution.
Figure\,\ref{fig:spec_xis} shows the XIS spectrum and best-fit models, whereas table\,\ref{tbl:param_fit_pin} summarizes the best-fit parameters.
We also fitted the XIS spectrum in the 0.5–-10\,keV band without fixing the plasma model, but the best-fit non-thermal parameters were consistent with each other within the statistical errors.

\par
\begin{figure}
  \begin{center}
    \includegraphics[width=85mm]{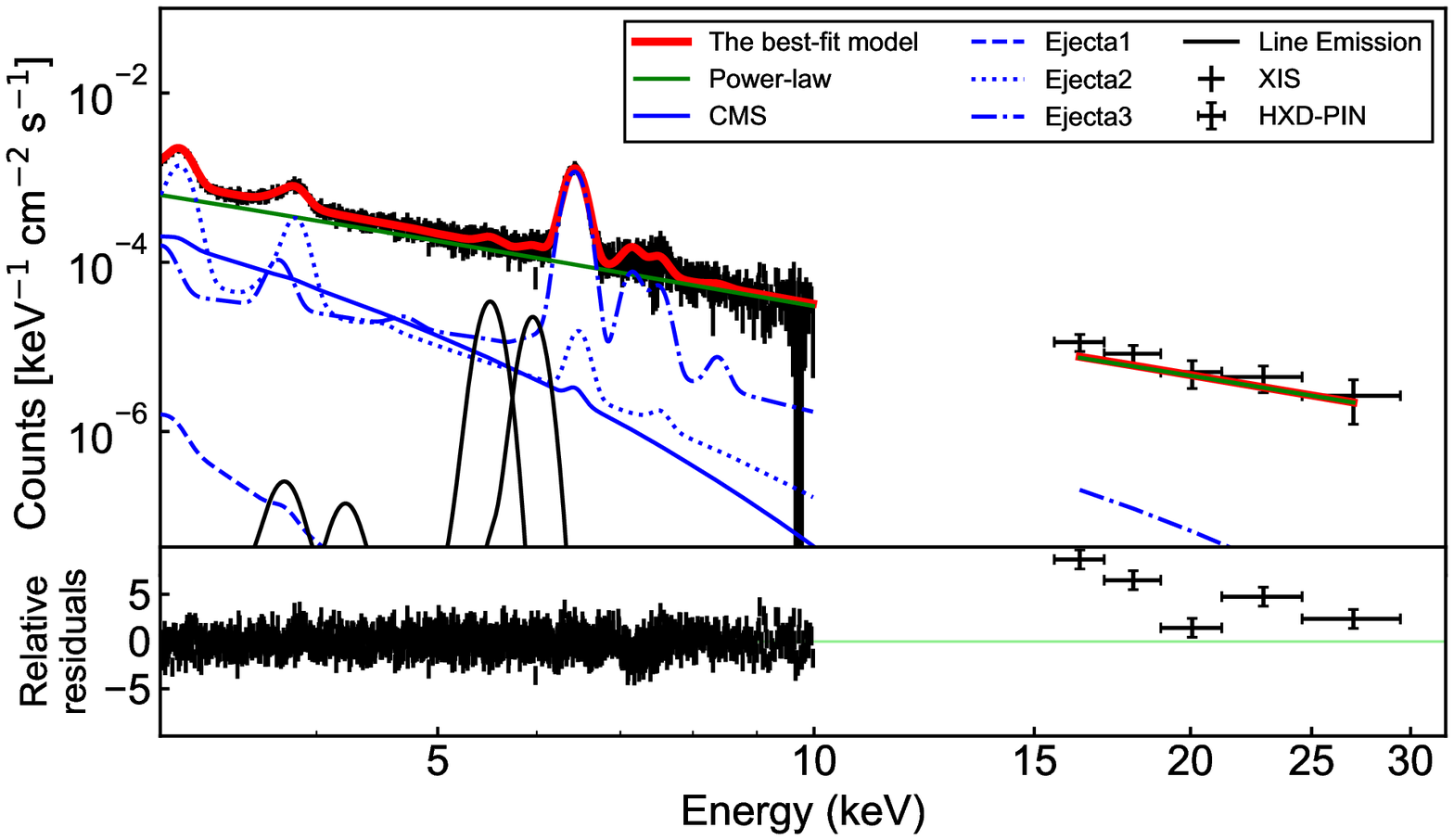}\\
    \includegraphics[width=85mm]{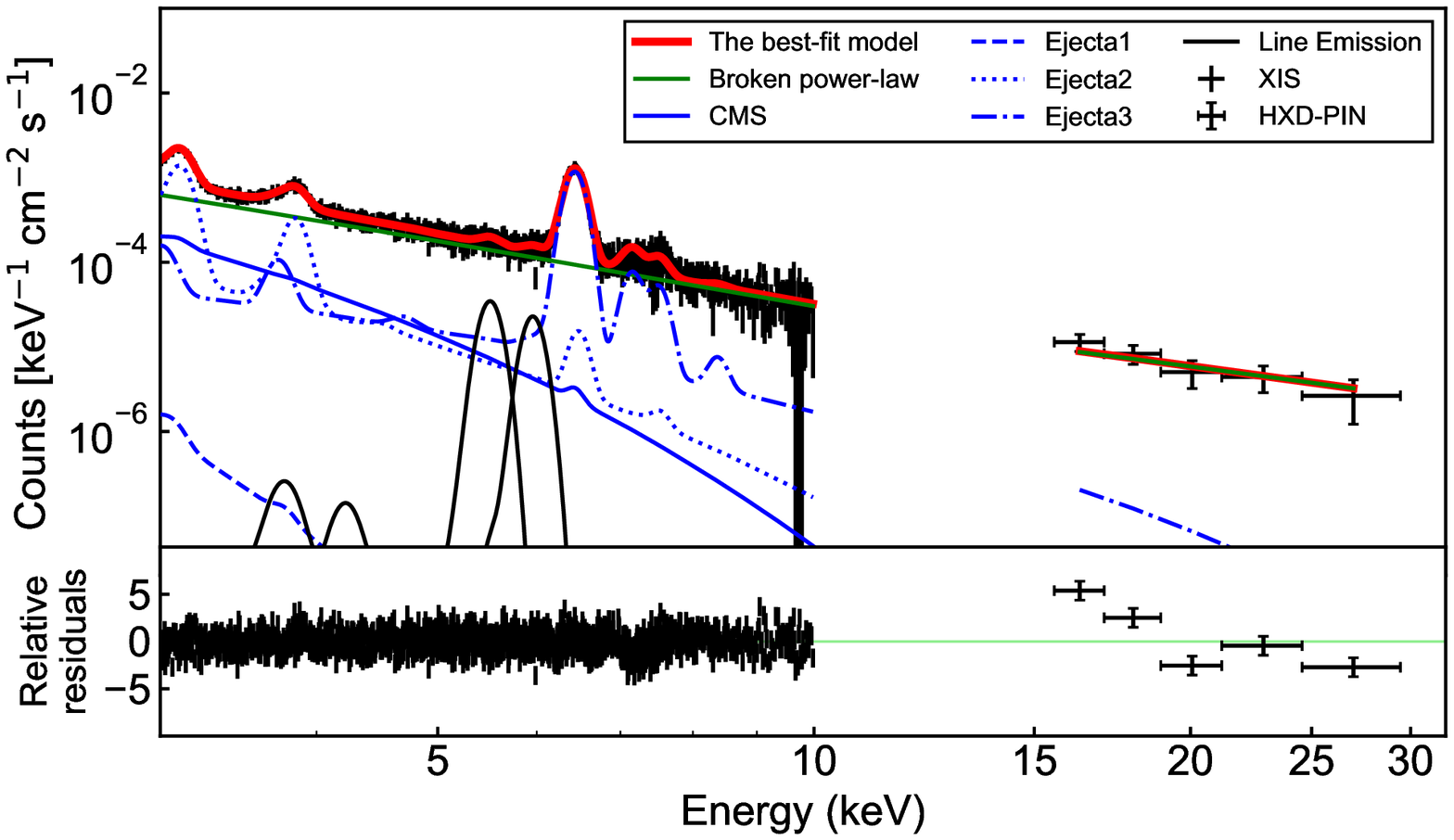}
  \end{center}
  \caption{Broad-band spectra of Kepler's SNR fitted with a model consisting of the thermal components described in \citet{2015ApJ...808...49K} and a non-thermal component of
    (Top panel)  single power-law  and (Bottom panel)  10-keV broken power-law. 
    The data points in the soft X-ray band\,(3--10\,keV)\, are taken from \citet{2015ApJ...808...49K}.}
  \label{fig:spec_broad_fit}
\end{figure}

The cross-normalization factor on the effective areas 
between the XIS and HXD-PIN for  a point-like source was estimated to be 1:1.15 \citep{2007PASJ...59S..53K}.
Using the correction factor of 0.96 for  the effective areas for the XIS and HXD spectra estimated in the previous subsection,
we carried out the model fitting for the 3--30\,keV band with the fixed cross-normalization of 1.15$\times$0.96$\sim$1.10.  
 As for the non-thermal component in the model,
we applied a single power-law  function (top panel of figure\,\ref{fig:spec_broad_fit}).
We also tested a broken power-law  function.  Our initial attempt yielded the best-fit broken energy  of 2.78\,keV, which is out of the energy band of the fitted spectra, with c-stat/dof=1.12.
 Thus, we fixed the broken energy at 10\,keV and refitted the spectrum (bottom panel of figure\,\ref{fig:spec_broad_fit}).
The best-fit parameters are  tabulated in the third and forth columns of table\,\ref{tbl:param_fit_pin}.
\par
\begin{figure}
  \begin{center}
    \includegraphics[width=85mm]{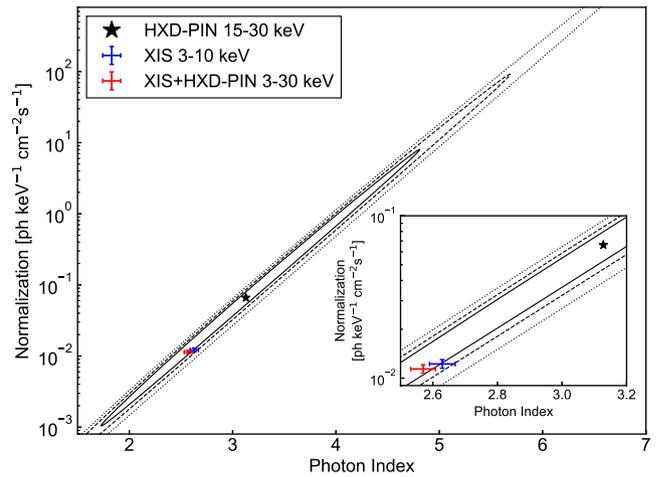}
  \end{center}
  \caption{
    Confidence contours between the photon index and normalization at 1\,keV  for a single power-law model.
    Solid, dashed, and dotted lines correspond to 1\,$\sigma$, 2\,$\sigma$, and 3\,$\sigma$ contours for the HXD-PIN spectrum, respectively.
    The black star  shows the best-fit value of HXD-PIN spectrum, whereas blue and red crosses show those
     of the XIS (3--10\,keV) and XIS+HXD-PIN (3--30\,keV), respectively.
  }
  \label{fig:pow_param}
\end{figure}

Furthermore, we investigated the parameter spaces of a power-low model, applying it to each spectrum separately, to see whether the photon-indices of the non-thermal emission between the soft X-ray (3--10\,keV) and hard X-ray (15--30\,keV) bands differ. 
Figure\,\ref{fig:pow_param} shows the confidence contour between the photon index and normalization at 1\,keV from single power-law models  applied individually to the HXD-PIN spectrum (15--30\,keV),
XIS spectrum (3--10\,keV), and combined XIS  and HXD-PIN spectra (3--30\,keV) where the cross-normalization factor for HXD-PIN data is taken into account.
 We found the best-fit power-law parameters  to be consistent in the 1-$\sigma$ confidence level  between one another.
\begin{table*}
  \tbl{Best-fit parameters for Kepler's SNR. Statistical errors are for the 90\% confidence.}
      {
        \renewcommand{\arraystretch}{1.2}
        \begin{tabular}{lcccc}
          \hline\noalign{\vskip3pt}
          Parameter & XIS & HXD-PIN\footnotemark[$\dagger$] & \multicolumn{2}{c}{XIS+HXD-PIN\footnotemark[$\dagger$]} \\
          \hline 
          Fitting range & 3-10\,keV & 15-30\,keV & 3-30\,keV & 3-30\,keV \\
          \hline\noalign{\vskip3pt}
          Thermal component & & & & \\
          CSM [$10^{10}\,$cm$^{-5}$]       & 344.80\footnotemark[$*$]   & - & 344.80\footnotemark[$*$] & 344.80\footnotemark[$*$] \\
          Ejecta\,1 [$10^{ 5}\,$cm$^{-5}$] & 961.42\footnotemark[$*$]   & - & 961.42\footnotemark[$*$] & 961.42\footnotemark[$*$] \\
          Ejecta\,2 [$10^{ 5}\,$cm$^{-5}$] & 2270.78$_{-50.96}^{+51.12}$ & - & 2263.27$\pm50.77\pm3.76$  & 2260.44$\pm50.85\pm0.14$  \\
          Ejecta\,3 [$10^{ 5}\,$cm$^{-5}$] & 2857.75$_{-36.1}^{+36.33}$  & - & 2831.10$\pm35.96\pm1.72$  & 2832.48$\pm35.99\pm0.05$  \\
          \hline\noalign{\vskip3pt}
          Non-thermal component & Single PL & Single PL & Single PL & Broken PL \\ 
          $\Gamma$(all or soft) & 2.63$_{-0.04}^{+0.04}$ & 3.13$_{-1.52-0.36}^{+1.85+0.69}$ & 2.57$\pm0.04^{+0.007}_{-0.008}$ & 2.57$\pm0.04^{+0.0001}_{-0.0004}$ \\
          $\Gamma$(hard)        & -                      & -                                & -                               & 2.10$\pm0.49\pm0.58$ \\
          Breaking energy [keV] & -                      & -                                & -                               & 10\footnotemark[$*$]  \\
          Flux\footnotemark[$\ddagger$] [$10^{-12}\,\mathrm{erg\,cm^{-2}\,s^{-1}}$] & 0.93$\pm0.15$ & 2.75$_{-0.77-0.82}^{+0.78+0.81}$ & 10.53$\pm0.16^{+0.03}_{-0.02}$ & 10.74$\pm0.26\pm0.25$ \\
          \hline\noalign{\vskip3pt}
          c-stat/dof & 1515.03/1411 & 0.37/3  & 1519.07/1416 & 1516.62/1415 \\
          \hline\noalign{\vskip3pt}
        \end{tabular}
      }
      \label{tbl:param_fit_pin}

      \begin{tabnote}
        \hangindent6pt\noindent
        \hbox to6pt{\footnotemark[$*$]\hss}\unskip: Fixed values.
      \end{tabnote}
      
      \begin{tabnote}
        \hangindent6pt\noindent
        \hbox to6pt{\footnotemark[$\dagger$]\hss}\unskip: The first and second errors are the statistical and systematic errors, respectively.
      \end{tabnote}
      
      \begin{tabnote}
        \hangindent6pt\noindent
        \hbox to6pt{\footnotemark[$\ddagger$]\hss}\unskip: The integration range is the same as the fitting range.
      \end{tabnote}      
\end{table*}

\section{Discussion and Summary}\label{sec:discussion}
Kepler's SNR has steep spectral indices ($\Gamma>2$) in both the soft and hard X-ray bands (\cite{2005ApJ...621..793B}, \cite{2015ApJ...808...49K}; see also section\,\ref{sec:analysis}).
The power-law component should represent the non-thermal emission.
Their origin can be explained by the synchrotron radiation or the non-thermal bremsstrahlung, as well known as non-thermal emissions from SNRs.
However, since the latter emission could be detected with the harder spectral index $\Gamma\sim1.4$ \citep{2018ApJ...866L..26T} than our result,
the former is more feasible to represent the XIS and HXD-PIN spectra by synchrotron emission at the highest energy of the electron population.
In this section,
we  model the broad-band non-thermal emission  from the radio  to TeV (upper limit) data and investigate the current particle distribution and properties of the local magnetic field.
\par
\begin{figure*}
  \begin{center}
    \includegraphics[width=170mm]{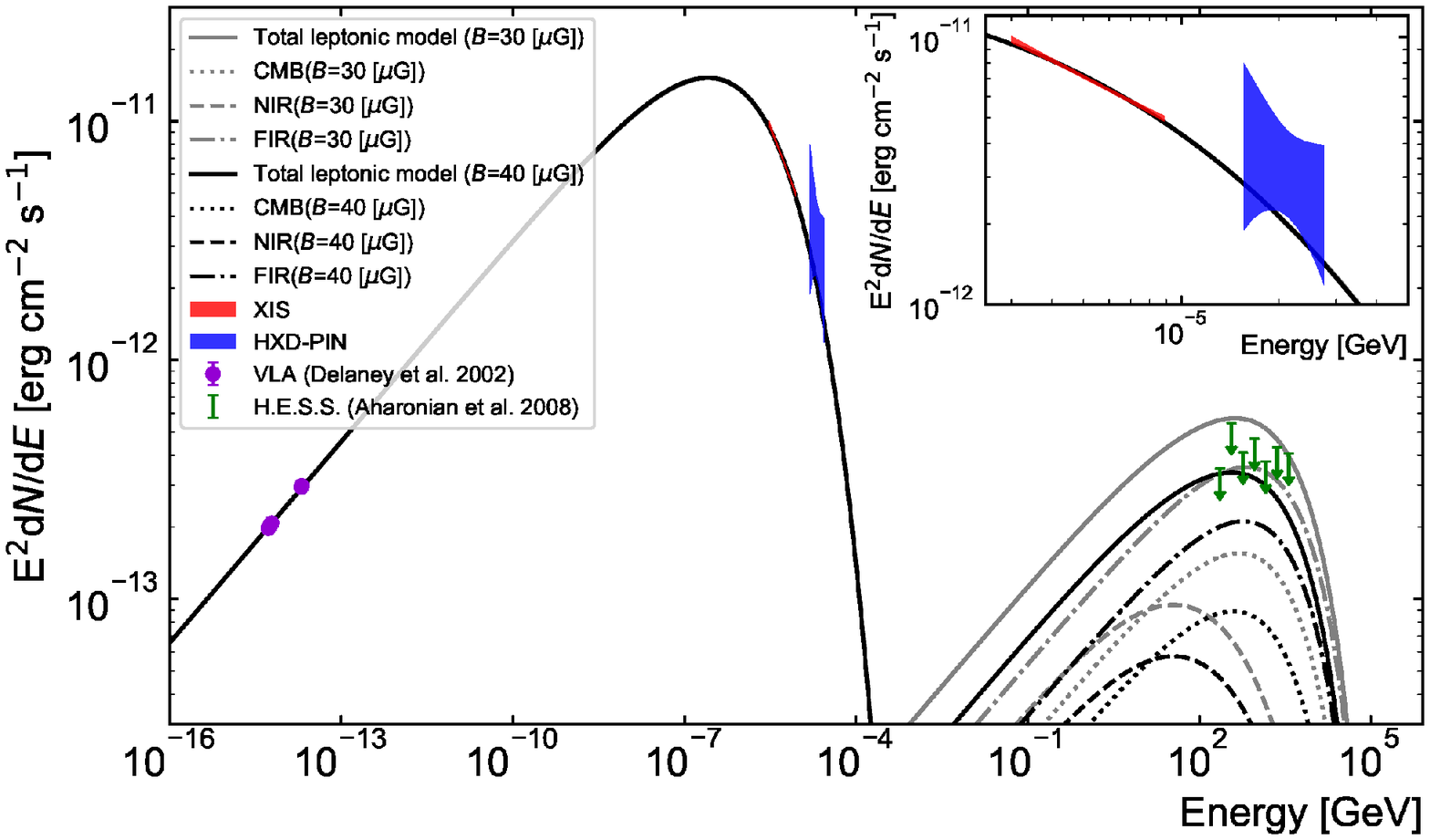}
  \end{center}
  \caption{Spectral energy distribution of the entire Kepler's SNR from the radio to TeV gamma-ray bands.
    The inset panel  shows zoomed-up data  in the X-ray band (1--30\,keV).
    The radio data points (magenta) are from \citet{2002ApJ...580..914D}.
    The TeV gamma-ray upper limits are from \citet{2008A&A...488..219A}.
    Red- and blue-shaded  regions are the non-thermal emission measured  with the XIS and HXD-PIN, respectively (see table\,\ref{tbl:param_fit_pin}).
    The red-shaded  region takes into account only the statistic error,  whereas the blue-shaded  region takes into account both the statistic and systematic errors.
  }
  \label{fig:sed}
\end{figure*}

Most of young SNRs are known to emit TeV gamma-ray emission
(e.g., \cite{2017ApJ...836...23A,2018A&A...612A...7H,2008AIPC.1085..304N,2018A&A...612A...6H,2008A&A...486..829A,2009ApJ...703L...6A,2017MNRAS.472.2956A}).
In the so-called leptonic model,  which is based on the assumption  that  their TeV gamma-ray emission is produced by the inverse Compton scattering  of high-energy electrons,
 the magnetic-field strength at an acceleration site can be estimated  with the equation $F_{\mathrm{TeV}}/F_{\mathrm{X}} \propto u_{\mathrm{rad}}/u_{\mathrm{B}}$,
where $F_{\mathrm{TeV}}$, $F_{\mathrm{X}}$, $u_{\mathrm{rad}}$, and $u_{\mathrm{B}}$ are the TeV flux, X-ray flux, energy density in the radiation field, and energy density of the magnetic filed, respectively.
The H.E.S.S. telescopes observed Kepler's SNR in 2004 and 2005 with a total live time of 13\,hrs and found no evidence for gamma-ray emission \citep{2008A&A...488..219A} with
 an estimated upper limit of  $8.6\times10^{-13}\,\mathrm{erg\,cm^{-2}\,s^{-1}}$ in  an energy range of 0.23--12.8\,TeV.
The upper limit gives a constraint on the high-energy particle distribution and magnetic field.  We use the radiative code and Markov Chain Monte Carlo (MCMC) fitting routines of {\it Naima} ver. 0.9.1 \footnote{http://naima.readthedocs.io/en/latest/index.html} to  estimate the present-age particle distribution \citep{2015ICRC...34..922Z} as follows.
First, we fit the radio and X-ray data with a simple model in which the radiating electrons are assumed to follow an exponential roll-off power-law distribution,
\begin{equation}
  N_\mathrm{e}\propto E_{\mathrm{e}}^{-p_{\mathrm{e}}}\exp{(-E_{\mathrm{e}}/E_{\mathrm{max,e}})},
  \label{eq:ecpl_simple}
\end{equation}
and obtain the amplitude of the electron distribution.
Then we calculate the flux of inverse Compton scattering (IC).
The seed photon fields considered for the IC emission are the cosmic microwave background (CMB) radiation,
a far-infrared (FIR) component with temperature $T=29.5\,\mathrm{K}$ and a density of 1.08$\,\mathrm{eV}\,\mathrm{cm^{-3}}$,
and a near-infrared (NIR) component with temperature $T=1800\,\mathrm{K}$ and a density of 2.25$\,\mathrm{eV}\,\mathrm{cm^{-3}}$.
The values for FIR and NIR are derived from GALPROP by \citet{2011ApJ...727...38S} for a distance of 4\,kpc.
Table\,\ref{tbl:electron_naima} lists the best-fit parameters, 
and figure\,\ref{fig:sed} shows the  obtained spectral energy distribution along with the radio and X-ray data and H.E.S.S. upper limits.
Note that the non-thermal emission measured by HXD-PIN (the blue-shaded region) shows slightly a higher flux than the model curve in figure\,\ref{fig:sed},
though they are consistent within the systematic error.
We confirmed that there is no hard X-ray sources in the field of view of the HXD-PIN (34$^\prime$ $\times$34$^\prime$)
by looking over the {\it INTEGRAL} catalog\footnote{http://www.isdc.unige.ch/integral/science/catalogue}.  
In order to further investigate a possible contamination of an additional emission, like non-thermal bremsstrahlung, 
we fitted the XIS $+$ HXD-PIN spectra with the model constructed in section\,\ref{subsec:broadfit} plus a second power-law component
with a fixed $\Gamma$ of 1.4 which is expected for the non-thermal bremsstrahlung (e.g., \cite{2018ApJ...866L..26T}).
Consequently, we obtained the c-value of 1519.08 which does not significantly improve from the original model.
Thus, the contamination from the non-thermal bremsstrahlung is not statistically significant
with the upper limit of the flux of 6.61$\times$10$^{-13}$ erg\,cm$^{-2}$\,s$^{-1}$ (68\% confidence range) in the 3--30\,keV band,
which is one order of magnitude lower than the flux of bremsstrahlung component of W49B \citep{2018ApJ...866L..26T}.
\begin{table}
  \tbl{The best-fit parameters of the radiating electron distribution  from the radio and X-ray data. The error represent 1$\sigma$ statistical uncertainty.}
      {%
        \begin{tabular}{cccc}
          \hline\hline
          $B$\,[$\mu$G] & $p_{\mathrm{e}}$\footnotemark[$*$] & $E_{\mathrm{max,e}}$\,[TeV]\footnotemark[$*$] & $W_{\mathrm{e}}$\,[erg]\footnotemark[$\dagger$] \\
          \hline\noalign{\vskip3pt}
          30 & 2.44$\pm0.01$ & 25.5$^{+0.7}_{-0.6}$ & 5.01$\pm0.04\times10^{47}$ \\
          40 & 2.44$\pm0.01$ & 21.9$^{+0.6}_{-0.7}$ & 3.04$\pm0.03\times10^{47}$ \\
          \hline\hline
        \end{tabular}
      }
      \label{tbl:electron_naima}
      \begin{tabnote}
        \hangindent6pt\noindent
        \hbox to6pt{\footnotemark[$*$]\hss}\unskip: $p_{\mathrm{e}}$ and $E_{\mathrm{max,e}}$ are  the spectral index and  maximum energy of the electron distribution defined in eq\,(\ref{eq:ecpl_simple}), respectively.
      \end{tabnote}
      
      \begin{tabnote}
        \hangindent6pt\noindent
        \hbox to6pt{\footnotemark[$\dagger$]\hss}\unskip: The total energy of the radiating electrons above 511\,keV  for an assumed distance of 4\,kpc.
      \end{tabnote}
\end{table}
\par
The best-fit spectral index $p_{\mathrm{e}}=2.44$ corresponds to the radio index $\alpha=-0.72$.
\citet{2002ApJ...580..914D} reported that  the radio spectral index $\alpha$ varied from place to place, ranging  from $-$0.85 to $-$0.60.
In the south-east region, where most of the X-ray emission is likely to be  non-thermal \citep{2004A&A...414..545C}, the radio spectral index  was distributed from $-$0.74 to $-$0.68 (see figure\,4. in  \cite{2002ApJ...580..914D}).
Under the assumption that  most of the non-thermal emission comes from the south-east region, the best-fit X-ray spectral index that we have obtained is consistent with that of the radio observation.
Consequently,  the averaged magnetic field strength of Kepler's SNR  is constrained to be $>$40\,$\mu$G from the H.E.S.S. upper limit.
This lower limit is consistent with the ones estimated  by \citet{2005ApJ...621..793B}.
Since the roll-off energy $\nu_{\rm roll}$ is determined  from the magnetic field strength ($B$) and  maximum electron energy $E_{\rm max,e}$, which satisfy
$\nu_{\mathrm{roll}} \sim B E_{\mathrm{max,e}}^2$ \citep{1999ApJ...525..368R},
we obtain
\begin{equation}
  \nu_{\mathrm{roll}} \simeq 1.0\times10^{17} \left(\frac{B}{40\,\mathrm{\mu G}}\right) \left(\frac{E_{\mathrm{max,e}}}{21.9\,\mathrm{TeV}}\right)^2 \,\mathrm{Hz}.
  \label{eq:emax_b}
\end{equation}
This $\nu_{\mathrm{roll}}$ is significantly smaller than the roll-off frequency
calculated from the spectra  of only the thin-filament structures by \citet{2005ApJ...621..793B}, $\sim3.6^{+3.3}_{-1.6}\times10^{17}$\,Hz.
The difference can be interpreted that magnetic field around the thin-filament region is damping (e.g., \cite{2005ApJ...626L.101P,2014ApJ...790...85R}) and
it results in higher roll-off energies at the filaments compared with a global one (cf. eq\,.(\ref{eq:emax_b})).
We estimate, from the (lower limit of) magnetic field strength of 40\,$\mu$G, the total energy of the radiating electron to be 3.04$\times$10$^{47}$\,erg at the distance of 4\,kpc.
Providing the proton and electron ratio $K_{\mathrm{ep}}=10^{-2}$ and conversion efficiency  for particle acceleration of 1\% ($\eta=0.01$),
the total energy of Kepler's SNR is calculated to be $\sim$10$^{51}$\,erg. It  is consistent with the standard energy of the SN explosion.
\par
 The spectral shape at around the maximum electron energy  provides key information about the acceleration mechanism of  electrons  in the SNR.
\citet{2014RAA....14..165Y} proposed a simple diagnostic method based on the observed synchrotron radiation in the soft and hard X-ray bands to constrain  the acceleration mechanism  in the SNR.
The energy spectrum of the electrons at around $E_{\mathrm{max,e}}$ follows a form of an exponential cutoff power-law
\begin{equation}
  N(E_{\mathrm{e}}) \propto E_{\mathrm{e}}^{-p_{\mathrm{e}}} \exp [(-E_{\mathrm{e}}/E_{\mathrm{max,e}})^{a}],
  \label{eq:ecpl_a}
\end{equation}
where $E_{\mathrm{e}}$, $p_{\mathrm{e}}$, and $a$ are the electron energy,  spectral index, and  cutoff shape parameter, respectively (see \citet{2014RAA....14..165Y}).
The simple diagnostic method presented  by \citet{2014RAA....14..165Y} uses the relation  between the soft and hard X-ray spectral indices.
\begin{figure}
  \begin{center}
    \includegraphics[width=85mm]{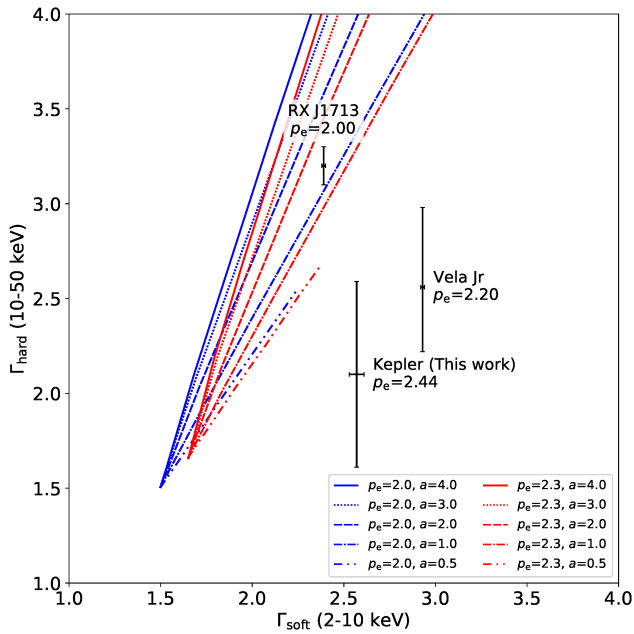}
  \end{center}
  \caption{
    Relation of the X-ray photon  indices  in the soft  and hard X-ray bands.
    The theoretical models derived by \citet{2014RAA....14..165Y} are  plotted in  lines in 2 colors (blue, red, and green)  in 5 styles to represent 2 and 5 parameter values for the spectral index $p_{\mathrm{e}}$ and cutoff shape parameter $a$ (see for detail \cite{2014RAA....14..165Y}).
    Black crosses show
    the measured photon indices of  three historic young SNRs: 
    RX\,J1713.7$-$3946 \citep{2008ApJ...685..988T},
    Vela\,Jr \citep{2016PASJ...68S..10T}, and
    Kepler's SNR (this work).
    The error bars represent  the 90\% confidence intervals.
  }
  \label{fig:diagnostics}
\end{figure}
\par
Figure\,\ref{fig:diagnostics} shows the predicted relation between the soft and hard X-ray indices  for 3 sets of $p_{\mathrm{e}}=2.0$, 3.0, and  2.3, which are taken from figures\,1 and 5 in \citet{2014RAA....14..165Y}.
We plot in the figure the  observed soft and hard X-ray spectral indices of  three young SNRs: RX\,J1713.7$-$3946 \citep{2008ApJ...685..988T}, Vela\,Jr \citep{2016PASJ...68S..10T}, and Kepler's SNR (this work).
Note that  the data point of Kepler's SNR from this work is taken from    the fourth column in table\,\ref{tbl:param_fit_pin} (see section~\ref{subsec:broadfit}).
We find that the data points of Vela\,Jr and Kepler’s SNR are off any of the predicted lines, even the most extreme case of $p_{\mathrm{e}}=3$ and $a\sim0.5$ in our set of examples (figure\,\ref{fig:diagnostics}).
The discrepancy may suggest that the particle acceleration is limited not  by synchrotron cooling but by their ages \citep{2014RAA....14..165Y}.
\par
The maximum electron energy in the age-limited case for Kepler's SNR is calculated to be, from eq\,(A.2) in \citet{2014RAA....14..165Y}, under an assumption of  the age of  400\,yr,
\begin{equation}
  E_{\mathrm{max,e}} \simeq 120
  \left(\frac{\eta}{1}\right)^{-1}
  \left(\frac{v_{\mathrm{s}}}{4,000\,\mathrm{km\,s^{-1}}}\right)^{2}
  \left(\frac{B}{40\,\mathrm{\mu G}}\right)
  \mathrm{TeV},
  \label{eq:emax_age_limit}
\end{equation}
where $v_{s}$ and $\eta$ are the shock speed and gyro-fractor, respectively.
Combining eqs\,(\ref{eq:emax_age_limit}) and (\ref{eq:emax_b}) yields the relation between $v_{\mathrm{s}}$ and $B$,
\begin{equation}
  v_{\mathrm{s}} \simeq 1,700
  \left(\frac{\eta}{1}\right)^{\frac{1}{2}}
  \left(\frac{B}{40\,\mathrm{\mu G}}\right)^{-\frac{3}{4}}
  \left(\frac{\nu_{\mathrm{roll}}}{1.0\times10^{17}\,\mathrm{Hz}}\right)^{\frac{1}{4}}
  \mathrm{km\,s^{-1}}.
  \label{eq:vs_final}
\end{equation}
In \citet{2020arXiv201201047T}, $\eta$ at the rims of Kepler's SNR are estimated $\eta$=0.3--3.2. Utilizing the values, $v_{\mathrm{s}}$ is estimated 930--3,000\,km\,s$^{-1}$.
This value is plausible, considering the measured shock speed at the rim of the remnant of 2,000--4,000\,km\,s$^{-1}$ \citep{2008ApJ...689..225K}.
At least, it does not exceed the highest speed of 10,400\,km\,s$^{-1}$ measured from the X-ray bright knots \citep{2017ApJ...845..167S}.
We conclude that the particle acceleration at Kepler's SNR is age-limited.
\par
In the case of age-limited acceleration, the synchrotron loss time $t_{\mathrm{sync}}$ should be larger than the age of Kepler's SNR ($t_{\mathrm{age}}\simeq$400\,yr).
From \citet{2012A&ARv..20...49V}, we estimate the synchrotron loss time adopting our modeling result, 
\begin{equation}
  t_{\mathrm{sync}}\simeq 640
  \left(\frac{\nu_{\mathrm{roll}}}{1.0\times10^{17}\,\mathrm{Hz}}\right)^{-\frac{1}{2}}
  \left(\frac{B}{40\,\mathrm{\mu G}}\right)^{-\frac{3}{2}}
  \,\mathrm{yr}.
  \label{eq:t_sync}
\end{equation}
Thus, the estimated $t_{\mathrm{sync}}$ is longer than $t_{\mathrm{age}}$, and
it indicates the obtained values are consistent with age-limited acceleration.
In addition, under the relation of $t_{\mathrm{age}}<t_{\mathrm{sync}}$, the upper limit of the magnetic field strength can be derived $B<$55\,$\mathrm{\mu G}$.
Taking into account the modeling result, the averaged magnetic field strength of Kepler's SNR is estimated $B=$40--55\,$\mathrm{\mu G}$.
Then, using eq\,(\ref{eq:emax_b}), the current maximum energy of electron is estimated $E_{\mathrm{max,e}}\simeq$19--22\,TeV.
Since the maximum energy of accelerated electrons is proportional to the remnant age under age-limited acceleration,
it is expected that Kepler's SNR will produce more energetic electrons.
It is also important to note that the age-limited acceleration would mean that Kepler's SNR accelerates protons up to the same energy as electrons. 
In order to study the detail of proton acceleration at SNR, GeV--TeV energy range is very helpful.
Due to low flux from Kepler's SNR in GeV--TeV range \citet{2016ApJS..224....8A,2008A&A...488..219A,2018A&A...612A...1H},
it is difficult to extract physical parameters regarding proton acceleration with the current GeV--TeV instrument.
This study could be performed by future CTA \citep{2011ExA....32..193A} observatory.
\begin{ack}
  We are grateful to the anonymous referee for providing us with valuable advice.
  We would like to thank all the members of the Suzaku team for their continuous efforts on developing and maintaining the data archive and performing instrument calibration  in Japan and the United States.
  We also thank Masaaki Sakano for helping to brush up this paper.
  This work was supported in part by JSPS KAKENHI Grant Number JP20K04009 (YT), 20H00174 (SK), 18H05459, No.19K03908 and Shiseido Female Researcher Science Grant (AB).
  This work was partly supported by Leading Initiative for Excellent Young Researchers, MEXT, Japan.
\end{ack}

\appendix

\appendix
\section{Thermal emission from Kepler's SNR}\label{sec:app_a}
The {\tt vnei} model  employed in the model-fitting for Kepler's SNR in this work  uses AtomDB\footnote{http://www.atomdb.org}.
AtomDB is an atomic database designed for X-ray plasma spectral modeling.
The current version of AtomDB is primarily used for modeling collisional plasma,  where hot electrons collide with (astrophysically abundant) elements and ions and  generate X-ray emission.
AtomDB  has been updated several times \footnote{http://www.atomdb.org/download.php}.
This  work uses AtomDB 3.0.9,  whereas \citet{2015ApJ...808...49K} used AtomDB 2.0.2.
 As a result, our fitting of an identical  dataset to those of \citet{2015ApJ...808...49K} has yielded  somewhat different results from theirs.
Table\,\ref{tbl:thermal_params_diff} summarizes the  fitting results  in the 0.5--7.0\,keV  band by \citet{2015ApJ...808...49K} and us.
Most of the best-fit parameters  between the two  works  agree within $\lesssim 30$\%.
\renewcommand{\arraystretch}{0.8}
\begin{table*}
  \tbl{The best-fit parameters for the thermal models of Kepler's SNR.}
      {
        \begin{tabular}{lccc}
          \hline\hline
          Parameter & \citet{2015ApJ...808...49K} & This work & Ratio\footnotemark[$*$] \\
          \hline
          CSM component & & & \\
          $kT_{\mathrm e}$ (keV) & 1.06$\pm$0.03 & 0.97$\pm$0.01 & 0.92$\pm$0.03 \\
          log($n_{\mathrm e}t$/cm$^{-3}\,$sec) & 10.81$^{+0.02}_{-0.01}$ & 10.71$_{-0.01}^{+0.01}$ & 0.99$\pm$0.00 \\
          Abundance$^a$ (solar)N & 3.31$^{+0.24}_{-0.25}$ & 1.93$_{-0.15}^{+0.17}$ & 0.58$_{-0.06}^{+0.07}$\\
          Redshift (10$^{-3}$) & 1.39$\pm$0.05 &  1.34$_{-0.04}^{+0.05}$ & 0.96$\pm$0.05 \\
          Line broadening (E/1\,keV eV) & 2.91$^{+0.36}_{-0.32}$ &  2.09$_{-0.22}^{+0.17}$ & 0.72$\pm$0.11 \\
          $\int n_{\mathrm e} n_{\mathrm H} dV$/4$\pi d^{2}$ ($10^{10}$cm$^{-5}$) & 193.15$^{+4.03}_{-6.48}$ & 344.80$_{-2.50}^{+3.98}$ & 1.79$_{-0.06}^{+0.04}$ \\
          \hline
          Ejecta components & & & \\
          (1)$kT_{\mathrm e}$ (keV) & 0.37$\pm$0.01 & 0.35$\pm0.00$ & 0.95$\pm$0.03 \\
          log($n_{\mathrm e}t$/cm$^{-3}\,$sec) & 10.52$\pm$0.01 & 10.75$_{-0.04}^{+0.02}$ & 1.02$\pm$0.00\\
          Abundance (10$^{4}$ solar)O & 0.25$^{+0.02}_{-0.03}$ & 0.05$_{-0.02}^{+0.02}$ & 0.20$\pm$0.08 \\
          Ne& 0.67$^{+0.06}_{-0.07}$ & 0.21$_{-0.03}^{+0.05}$ & 0.31$_{-0.06}^{+0.08}$ \\
          Mg& 0.77$^{+0.07}_{-0.06}$ & 0.48$\pm0.05$          & 0.62$_{-0.08}^{+0.09}$ \\
          S& 15.53$^{+0.22}_{-0.14}$ & 11.20$_{-0.07}^{+0.09}$ & 0.72$\pm$0.01 \\
          Ar& 18.98$^{+0.46}_{-0.52}$& 16.74$_{-0.45}^{+0.42}$ & 0.88$\pm$0.03 \\
          Ca& 36.40$^{+1.51}_{-1.60}$& 31.43$_{-1.37}^{+1.16}$ & 0.86$\pm$0.05 \\
          Fe& 13.65$^{+0.17}_{-0.47}$& 14.29$_{-1.79}^{+1.06}$ & 1.05$_{-0.14}^{+0.08}$ \\
          $\int n_{\mathrm e} n_{\mathrm H} dV$/4$\pi d^{2}$ ($10^{5}$cm$^{-5}$) & 961.42$^{+114.02}_{-9.8}$ & 1843.48$_{-12.31}^{+20.87}$ & 1.92$_{-0.02}^{+0.23}$\\
          Redshift (10$^{-3}$) & -2.94$\pm$0.01 & -0.81$_{-0.01}^{+0.01}$ & 0.28$\pm$0.00 \\
          Line broadening (E/1\,keV eV) & 9.13$^{+0.14}_{-0.21}$ & 5.48$_{-0.14}^{+0.09}$ & 0.60$_{-0.02}^{+0.01}$ \\
          (2)$kT_{\mathrm e}$ (keV) & 2.08$^{+0.01}_{-0.02}$ & 1.435$\pm0.00$ & 0.69$_{-0.01}^{+0.00}$ \\
          log($n_{\mathrm e}t$/cm$^{-3}\,$sec) & 10.32$\pm$0.01 & 10.47$_{-0.00}^{+0.01}$ & 1.01$\pm$0.00 \\
          Abundance$^c$ (10$^{4}$ solar)Fe & 3.58$^{+0.04}_{-0.05}$ & 2.57$_{-0.08}^{+0.02}$ & 0.72$_{-0.02}^{+0.01}$ \\
          $\int n_{\mathrm e} n_{\mathrm H} dV$/4$\pi d^{2}$ ($10^{5}$cm$^{-5}$) & 875.35$^{+4.73}_{-4.19}$ & 2228.54$_{-10.33}^{+9.52}$ & 2.55$\pm$0.02 \\
          (3)$kT_{\mathrm e}$ (keV) & 2.59$\pm$0.01 & 3.74$_{-0.03}^{+0.12}$ & 1.44$_{-0.01}^{+0.05}$ \\
          log($n_{\mathrm e}t$/cm$^{-3}\,$sec) & 9.21$^{+0}_{-0}$ & 9.34$_{-0.00}^{+0.02}$ & 1.01$\pm0.00$ \\
          Abundance$^d$ (10$^{4}$ solar)Ar & 0 ($<0.19$) & 2.47$_{-0.53}^{+0.59}$ & - \\
          Ca & 1.43$^{+0.31}_{-0.28}$ & 4.24$_{-0.79}^{+0.73}$ & 2.97$_{-0.80}^{+0.82}$ \\
          Redshift (10$^{-3}$) & -5.73$^{+0.14}_{-0.18}$ & -4.46$_{-0.10}^{+0.30}$ & 0.78$_{-0.03}^{+0.06}$ \\
          Line broadening (E/1\,keV eV) & 12.10$\pm$0.27 & 7.486$_{-0.16}^{+0.16}$ & 0.62$\pm$0.02 \\
          $\int n_{\mathrm e} n_{\mathrm H} dV$/4$\pi d^{2}$ ($10^{5}$cm$^{-5}$) & 5808.47$^{+71.11}_{-63.17}$ & 2823.47$_{-34.05}^{+30.07}$ & 0.49$\pm$0.01 \\
          \hline
          Additional lines & & & \\
          FeL+OKCenter (keV) & 0.708$^{+0.002}_{-0.001}$ &  0.72$_{-0.001}^{+0.00}$ & 1.02$\pm$0.00 \\
          Norm (10$^{-4}$ ph\,keV$^{-1}$\,cm$^{-2}$\,s$^{-1}$) & 68.18$^{+5.4}_{-5.0}$ & 196.07$_{-10.67}^{+8.96}$ & 2.88$\pm$0.26 \\
          FeL+NeKCenter (keV) & 1.227$^{+0.001}_{-0.003}$ & 1.227$_{-0.003}^{+0.001}$ & 1.02$\pm$0.00 \\
          Norm (10$^{-4}$ ph\,keV$^{-1}$\,cm$^{-2}$\,s$^{-1}$) & 22.75$^{+1.2}_{-1}$ & 23.66$_{-0.10}^{+1.20}$ & 1.04$_{-0.05}^{+0.08}$\\
          CrKCenter (keV) & 5.514$^{+0.035}_{-0.037}$ & 5.51$_{-0.03}^{+0.03}$ & 1.00$\pm$0.01 \\
          Norm (10$^{-7}$ ph\,keV$^{-1}$\,cm$^{-2}$\,s$^{-1}$) & 85.18$^{+16.7}_{-17.1}$ & 0.10$\pm0.02$ & (1.17$\pm$0.33)$\times10^{-3}$ \\
          MnKCenter (keV) & 5.976$^{+0.04}_{-0.042}$ & 5.960$_{-0.034}^{+0.041}$ & 1.00$\pm$0.01 \\
          Norm (10$^{-7}$ ph\,keV$^{-1}$\,cm$^{-2}$\,s$^{-1}$) & 60.18$\pm$16.3 & 0.07$\pm0.02$ & (1.16$\pm$0.46)$\times10^{-3}$ \\
          \hline
        \end{tabular}
      }
      \label{tbl:thermal_params_diff}
      \begin{tabnote}
        \hangindent6pt\noindent
        \hbox to6pt{\footnotemark[$*$]\hss}\unskip: The ratio is the values obtained  in this work  to  those obtained by \citet{2015ApJ...808...49K}.
      \end{tabnote}
\end{table*}

\end{document}